\documentclass[amsmath,12pt]{article}
\usepackage[dvips]{epsfig}

\begin{document}

\title{\huge \bf Fourth generation effects in the $B_s \to \nu \bar{\nu} \gamma$ decay}
\author{Yusuf Din\c{c}er\footnote{e-mail:dincer@physik.rwth-aachen.de} \\
        Institute of Theoretical Physics, RWTH Aachen \\
        D-52056 Aachen, Germany}
\date{}
\maketitle

\begin{abstract}

If the fourth generation fermions exist, the new quarks could influence the branching ratio of the decay $B_s \to \nu \bar{\nu} \gamma$. We obtain two solutions of the fourth
generation CKM factor $V^*_{t's} V_{t'b}$ from the decay of $B \to X_s \gamma$. With these two solutions we calculate the new contributions of the fourth generation quark 
to Wilson coefficients of the decay $B_s \to \nu \bar{\nu} \gamma$. The branching ratio of the decay $B_s \to \nu \bar{\nu} \gamma$ in the two cases are calculated. In one
case, our results are quite different from that of SM, but almost same in another case. If a fourth generation should exist in nature and nature chooses the former case, this
B meson decay could provide a possible test of the fourth generation existence.

\end{abstract}

\newpage

\section{\huge Introduction}

At present the Standard Model (SM) describes very succesfully all low energy experimental data. But there is no doubt that from a theoretical point of view SM is an 
incomplete theory. Among the unsolved problems, such as CP violation, mass spectrum, etc., one of the unsolved mysteries of the SM is the number of generations. In SM
there are three generations and, yet, there is no theoretical argument to explain why there are three and \textsl{only} three generations in SM. From the LEP result of the 
invisible partial decay width of the Z boson it follows that the mass of the extra generation neutrino N should be larger than $45$ GeV \cite{Collab}. 
There is neither an experimental evidence for a fourth generation nor does any experiment exclude such extra generations. Having this 
experimental result in mind we can raise the following question: If extra generations really exist, what effect would they have in low energy physics?

In \cite{Huo} effects of the fourth generation quarks on $\Delta M_{B_{d,s}}$ in $B^0-\bar{B}^0$ mixing is discussed. In \cite{Ginzburg}  
it is argued that the fourth generation quarks and
leptons can manifest themselves in the Higgs boson production at the Tevatron and the LHC, before being actually observed. The next generation leads to an increase of the 
Higgs boson production cross section via gluon fusion at hadron colliders by a factor $6-9$. So, the study of this process at the Tevatron and LHC can fix
the number of generations in the SM. In \cite{Gunion} 
the possibility of a fourth generation in the Minimal Supersymmetric Standard Model (MSSM) is explored. It is shown that the
new generations must have masses $m_{\nu'},m_{\tau'} < 86$ GeV, $m_{t'}<178$ GeV and $m_{b'}<156$ GeV so that the MSSM remains perturbative up to the unification scale $M_U$ of
the Yukawa couplings. In \cite{Froggatt} 
even the possibility of the fourth generation of quarks without the fourth generation leptons is discussed. In \cite{Beneke} the decay of the top quark 
into a possible fourth generation $b'$ is regarded. The effect of the fourth generation on the branching ratio of the $B \to X_s l^+ l^-$ and the $B \to X_s \gamma$ decays is
analysed in \cite{Huang}. 
It is also shown that the fourth generation could influence the forward-backward asymmetry of the decay of $B \to X_s l^+ l^-$. 
The introduction of fourth generation fermions can also affect CP violating parameters $\epsilon' / \epsilon$
in the Kaon system \cite{Chou}. In \cite{Aliev3} the fourth generation effects in the rare exclusive $B \to K^* l^+ l^-$ decay are studied.

In \cite{Evans}-\cite{Maltoni}, 
contributions of the new generation to the electroweak radiative corrections were considered. It was shown in \cite{Maltoni} that the existing electroweak data on the 
Z-boson parameters, the W-boson and the top quark masses strongly excluded the existence of the new generations with all fermions heavier than the Z boson mass. However the 
same allows few extra generations, if one allows neutral leptons to have masses close to $50$ GeV.

One promising area in experimental search of the fourth generation, via its indirect loop effects, are the B meson decays. It is hoped that we will find a definite answer on a 
possible fourth generation at the B factories at SLAC and KEK.

In this work we study the contribution of the fourth generation in the $B_s \to \nu \bar{\nu} \gamma$ decay. New physical effects can manifest themselves through the Wilson 
coefficients, whose values can be different from the ones in  the SM \cite{Goto,Huang}, 
as well as through new operators \cite{Aliev2}. But in this work, we will only regard the contribution of 
the fourth generation in the Wilson coefficients and assume that the operators in SM for three (SM) and four generations (SM4) are the same.
For the form factors we use the results from the Light Cone QCD sum rule methods \cite{Aliev2}. 
We use the dipole formula approximation for the form factors which are known with an
uncertainty of about $20 - 30 \%$ \cite{Aliev2}. However, it will be possible to reduce this uncertainty if the decay $B_s \to \mu \nu \gamma$ is detected.
Hence, experimental deviations from the SM prediction on the branching ratio for the $B_S \to \nu \bar{\nu} \gamma$ decay of about
$20 -30 \%$ will not necessarily be a signal for new physics. But deviations of more than $30 \%$ could be a signal for new physics. And one possiblity for new physics
could be the extension of the SM to four generations.

The paper is organized as follows. In section \ref{effective} we establish the effective Hamiltonian describing this decay. 
In section \ref{numerical} we present the numerical values. Finally, in section \ref{conclusion} we summarize the results and give an outlook.


\section{\huge The effective Hamiltonian}\label{effective}

The weak decay of mesons is described by the effective Hamiltonian
\begin{equation}
{\cal H}_{\rm eff} = \frac{4 G_F}{\sqrt{2}} V_{tb} V^*_{ts} \sum_{i=1}^{10} C_i(\mu) {\cal O}_i(\mu) \,,
\end{equation}
where the full set of the operators ${\cal O}_i(\mu)$ and the corresponding expressions for the Wilson coefficients $C_i(\mu)$ in the SM are given in \cite{Buras}. 
As we mentioned in
the introduction, no new operators appear and clearly the full operator set is exactly the same as in SM. The fourth generation changes only the values of the Wilson
coefficients $C_7(\mu),C_9(\mu)$ and $C_{10}(\mu)$, via virtual exchange of the fourth generation up quark $t'$. The above mentioned Wilson coefficients can be written in the 
following form
\begin{eqnarray}\label{wilson}
C^{\rm SM4}_7 (\mu)  &=& C^{\rm SM}_7(\mu) + \frac{ V^*_{t'b} V_{t's}}{V^*_{tb} V_{ts}} C^{\rm new}_7(\mu) \,, \\
C^{\rm SM4}_9 (\mu) &=& C^{\rm SM}_9(\mu) + \frac{ V^*_{t'b} V_{t's}}{V^*_{tb} V_{ts}} C^{\rm new}_9(\mu) \,, \\
C^{\rm SM4}_{10} (\mu) &=& C^{\rm SM}_{10}(\mu) + \frac{ V^*_{t'b} V_{t's}}{V^*_{tb} V_{ts}} C^{\rm new}_{10}(\mu) \,, \label{wilson2}
\end{eqnarray}
where the last terms in these expressions describe the contributions of the $t'$ quark to the Wilson coefficients and $V_{t'b}$ and $V_{t's}$ are the two elements of the
$(4 \times 4)-$CKM matrix. If the quark $b'$ should be very massive then it will also give an additional contribution to the Wilson coefficients on the right hand side of
the Eqs. (\ref{wilson})-(\ref{wilson2}). But we neglect such a contribution in this paper.
The explicit forms of the $C^{\rm new}_7,C^{\rm new}_9$ and $C^{\rm new}_{10}$ can easily be obtained from the corresponding Wilson coefficient
expressions in SM by simply substituting $m_t \to m_{t'}$ \cite{Buras,Grinstein}.
Here, the effective Hamiltonian for the $B_s \to  \nu \bar{\nu} \gamma$ decay in the SM4 is given by
\begin{equation}
{\cal H}_{\rm eff} = C^{\rm SM4} [ \bar{s} \gamma_\mu(1-\gamma_5) b] \cdot [ \bar{\nu} \gamma^\mu (1-\gamma_5) \nu ] \,,
\end{equation}
where
\begin{equation}
C^{\rm SM4} = C^{\rm SM} + \frac{ V^*_{t'b} V_{t's}}{V^*_{tb} V_{ts}} \, C^{\rm new}
\end{equation}
and
\begin{equation}\label{sminami}
C^{\rm SM} = \frac{G_F \alpha}{2 \sqrt{2} \pi \sin^2 \Theta_W} V_{tb} V^*_{ts} \frac{x}{8}
\left[ \frac{x+2}{x-1} + \frac{3(x-2)}{(x-1)^2} \ln x \right] \,, \;  x = (m_t/m_W)^2 \,.
\end{equation}
As mentioned, we obtain $C^{\rm new}$ form $C^{\rm SM}$ (\ref{sminami}) by substituting $m_t \to m_{t'}$. 
The corresponding matrix element for the process $B_s \to \nu \bar{\nu} \gamma$ is given by
\begin{equation}
{\cal M} = C^{\rm SM4} [ \bar{\nu}(p_2) \gamma_\mu (1-\gamma_5) \nu(p_1) ] < \gamma(q) \vert \bar{s} \gamma^\mu (1-\gamma_5) b \vert B(q+p) > \,,
\end{equation}
where $q^2 = (p_1+p_2)^2$ and $p_1$ and $p_2$ are the final neutrinos four momenta. The matrix element 
$< \gamma(q) \vert \bar{s} \gamma^\mu (1-\gamma_5) b \vert B(q+p) >$ can be parametrized in terms of the two gauge invariant and independent form factors $f(p^2)$ and 
$g(p^2)$,namely
\begin{equation}
< \gamma \vert \bar{s} \gamma^\mu (1-\gamma_5) b \vert B > = \sqrt{4 \pi \alpha} \left[ \epsilon_{\mu \alpha \beta \sigma} \varepsilon^*_\alpha p_\beta q_\sigma 
\frac{g(p^2)}{m^2_B} + i \left( \varepsilon^{*}_\mu (pq) - (\varepsilon^{*} p)q_\mu \right) \frac{f(p^2)}{m^2_B} \right] \,. \nonumber
\end{equation}
Here, $\varepsilon_\mu$ and $q_\mu$ stand for the polarization vector and momentum of the photon, $p+q$ is the momentum of the B meson, $g(p^2)$ and $f(p^2)$ correspond
to parity conserved and parity violated form factors for the $B_s \to \nu \bar{\nu} \gamma$ decay. 
The form factors $f(p^2)$ and $g(p^2)$ were calculated in \cite{Aliev2} with the light cone QCD sum rules method.
As mentioned in \cite{Aliev2}, the best agreement is achieved with the dipole formulae 
\begin{equation}
g(p^2) \approx \frac{h_1}{(11-\frac{p^2}{m_1^2})^2} \qquad \,,  \qquad f(p^2) \approx \frac{h_2}{(1-\frac{p^2}{m_2^2})^2}
\end{equation}
with
$h_1 \approx 1.0$ GeV  ,  $m_1 \approx 5.6$ GeV   ,  $h_2 \approx 0.8$ GeV     and    $m_2 \approx 6.5$ GeV \,.
For the total decay rate we get
\begin{equation}
\Gamma ( B_s \to \nu \bar{\nu} \gamma ) = \frac{\alpha (C^{\rm SM4})^2 m^5_B}{256 \pi}    \,     I      \,,
\end{equation}
where
\begin{equation}
I = \frac{1}{m^2_B}  \int_0^1 dx \, (1-x)^3 \, x \, \left( f^2(x) + g^2(x) \right) \,.
\end{equation}
Here $x=1-\frac{2E_\gamma}{m_B}$ is the normalized photon energy.

In order to obtain quantitative results we need the value of the fourth generation CKM matrix element $\vert V^*_{t's} V_{t'b} \vert$. Following \cite{Huang}, we will use the 
experimental results of the decays $Br(B \to X_s \gamma)$ and $Br(B \to X_c e \bar{v}_e)$ to determine the fourth generation CKM factor $V^*_{t's} V_{t'b}$. In order to 
reduce the uncertainties arising from $b$ quark mass, we consider the following ratio
\begin{equation}
R = \frac{Br(B \to X_s \gamma)}{Br(B \to X_c e \bar{v}_e)} \,.
\end{equation}
In leading logarithmic approximation this ratio can be written as
\begin{equation}\label{r}
R = \frac{ \vert V^*_{ts} V_{tb} \vert^2}{\vert V_{cb} \vert^2} \frac{6 \alpha \vert C^{\rm SM4}_7(m_b) \vert^2}{\pi f(\hat{m}_c) \kappa(\hat{m}_c)} \,,
\end{equation}
where the phase factor $f(\hat{m}_c)$ and ${\cal O}(\alpha_s)$ QCD correction factor $\kappa(\hat{m}_c)$ \cite{Buras2} of $b \to c l \bar{\nu}$ are given by
\begin{eqnarray}\label{fkappa}
f(\hat{m}_c) &=& 1 - 8 \hat{m}_c^2 + 8 \hat{m}^6_c - \hat{m}^8_c - 24 \hat{m}_c^4 \ln ( \hat{m}^4_c) \,, \\
\kappa( \hat{m}_c) &=& 1 - \frac{2 \alpha_s(m_b)}{3 \pi} \left[ \left( \pi^2 - \frac{31}{4} \right) (1-\hat{m}_c)^2 + \frac{3}{2} \right] \,.
\end{eqnarray}
Solving Eq. (\ref{r}) for $V^*_{t's} V_{t'b}$ and taking into account (\ref{wilson}) and (\ref{fkappa}), we get
\begin{equation}
V^*_{t' s} V_{t' b}^\pm = \left[ \pm \sqrt{ \frac{\pi R \vert V_{cb} \vert^2 f( \hat{m}_c ) \kappa( \hat{m}_c )}{6 \alpha \vert V^{*}_{ts} V_{tb} \vert^2} } -
C_7^{\rm SM}(m_b) \right] \frac{V^{*}_{ts} V_{tb}}{C_7^{\rm new}(m_b)} \,.
\end{equation}
The values for $V^*_{t's} V^\pm_{t'b}$ are listed in Table \ref{tabelle1} and \ref{tabelle2}.

\begin{figure}
\epsfig{file=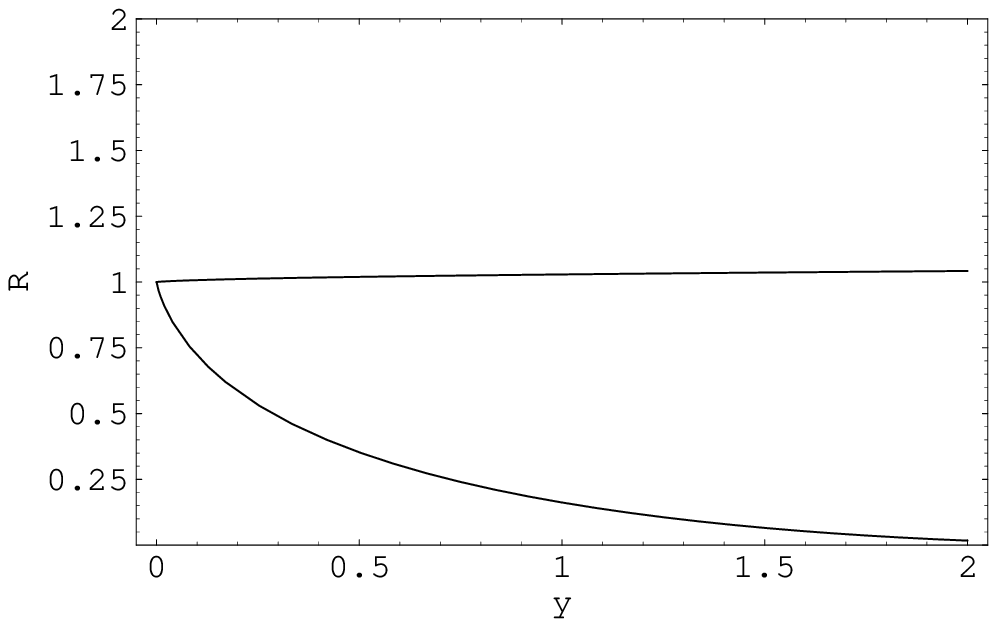}
\caption{The ratio $R$ versus $y=(m_{t'}/m_W)^2$;
the upper curve is for the solution $V^*_{t's} V^{(-)}_{t'b}$; the lower curve is for the solution $V^*_{t's} V^{(-)}_{t'b}$.}
\label{figur}
\end{figure}

From unitarity condition of the CKM matrix we have
\begin{equation}\label{unitarity}
V^*_{us} V_{ub} + V^*_{cs} V_{cb} + V^*_{ts} V_{tb} + V^*_{t's} V_{t'b} = 0 \,.
\end{equation}
If the average values of the CKM matrix elements in the SM are used, the sum of the first three terms in Eq. (\ref{unitarity}) is about $7.6 \times 10^{-2}$. 
Substituting the value of $V^*_{t's} V^{(+)}_{t'b}$ from Table \ref{tabelle1} and \ref{tabelle2}, 
we observe that the sum of the four terms on the left-hand side of Eq. (\ref{unitarity}) is closer to zero compared to
 the SM case, since $V^*_{t's} V_{t'b}$ is very close to the sum of the first three terms, but with opposite sign. On the other hand if we consider $V^*_{t's} V^{(-)}_{t'b}$,
whose value is about $10^{-3}$ and one order of magnitude smaller compared to the previous case. However it should be noted that the data for the CKM is not determined to a
very high accuracy, and the error in sum of first three terms in Eq. (\ref{unitarity}) is about $\pm 0.6 \times 10^{-2}$. It is easy to see then that the value of 
$V^*_{t's} V^{(-)}_{t'b}$ is within this error range. In summary both $V^*_{t's} V_{t'b}^{(+)}$ and $V^*_{t's} V^{(-)}_{t'b}$ satisfy the unitarity condition 
(\ref{unitarity}) of CKM. 
Moreover, since $\vert V^*_{t's} V^{(-)}_{t'b} \vert \le 10^{-1} \times \vert V^*_{t's} V^{(+)}_{t'b} \vert$, $V^*_{t's} V^{(-)}_{t'b}$ contribution to the physical quantities
should be practically indistinguishable from SM results, and our numerical analysis confirms this expectation. This can also be seen in Figure \ref{figur}. There, the
quantity $R = \Gamma ( B_s \to \nu \bar{\nu} \gamma )_{\rm SM4} /  \Gamma ( B_s \to \nu \bar{\nu} \gamma )_{\rm SM}$ is plotted as a function of $y=(m_{t'}/m_W)^2$. For
$V^*_{t's} V^{(-)}_{t'b}$ this ratio $R$ is approximately one.  The greater $m_{t'}$ is the more the ratio $R$ differs from unity.


\section{\huge Numerical Analysis}\label{numerical}

In this section we will calculate the branching ratio in SM4 and study the influence of the fourth generation to the branching ratio. In \cite{Aliev2} 
the branching ratio in SM with three generations was calculated to 
\begin{equation}\label{smresult}
Br ( B_s \to \nu \bar{\nu} \gamma ) \approx 7.5 \times 10^{-8} \,.
\end{equation}
We use the following numerical values
\begin{eqnarray*}
\alpha &=& 1/137 \,, e = \sqrt{4 \pi \alpha} \,,      \sin^2 \Theta_W = 0.2319 \,, G_F = 1.16639 \times 10^{-5} \, {\rm GeV^{-2}} \,, \\
m_W &=& 80.22 \, {\rm GeV} \,, m_b = 4.8 \, {\rm GeV} \,, m_t = 176 \, {\rm GeV} \,, m_s = 0.51 \, {\rm GeV} \,, m_{B_s} = 5.3 \, {\rm GeV} \,, \\
m_c &=& 1.6 \, {\rm GeV} \,, m_d = 0.25 \, {\rm GeV} \,,  \vert V^*_{ts} V_{tb} \vert = 0.045 \,.
\end{eqnarray*}
For the values $V^*_{t's} V^{(\pm)}_{t'b}$ see Table \ref{tabelle1} and \ref{tabelle2}. 
The results on the branching ratios in SM4 are given in the Table \ref{tabelle1} and \ref{tabelle2}. For the 
$B_s \to \nu \bar{\nu} \gamma$ decay for both solutions $V^*_{t's} V^{(\pm)}_{t'b}$ we see that for the choice $V^*_{t's} V^{(-)}_{t'b}$ the branching ratios for the
decay $B_s \to \nu \bar{\nu} \gamma$ calculated in SM4 coincide with the result (\ref{smresult}) from SM. 
However, when we choose $V^*_{t's} V^{(+)}_{t'b}$, we observe 
significant deviations from the SM. We observe that the branching ratio in SM4 is smaller than in SM for values $m_{t'} < m_t$. But it increases for enlarging the mass
$m_{t'}$.


\section{\huge Conclusion}\label{conclusion}

In this work, we have studied the decay process $B_s \to \nu \bar{\nu} \gamma$ in the Standard Model with four generations. We obtained two solutions of the fourth
generation CKM factor $V^*_{t's} V_{t'b}$ from the experimental data of $B \to X_s \gamma$. We have used the two solutions to calculate the contributions of the fourth 
generation quark $t'$ to the Wilson coefficients. We have calculated the branching ratio for this process in the two cases and compared our 
results with those from the SM with three generations. We found that the new results are different from that of SM when the value of the fourth generation CKM factor is
negative, but almost the same when the value is positive. Therefore, the decay $B_s \to \nu \bar{\nu} \gamma$ could provide a possible way to probe the existence of the 
fourth generation if the fourth generation CKM factor $V^*_{t's} V_{t'b}$ is negative.


\section{\huge Acknowledgement}

The author wants to thank Prof. L.M. Sehgal for helpful suggestions.

\newpage

\begin{table}
\begin{tabular}{||c|c|c||}
\hline
$m_{t'} [GeV]$ & $V^*_{t's} V^{(-)}_{t'b} \times 10^{-3}$ & $Br(B_s \to \nu \bar{\nu} \gamma) [10^{-8}]$ \\
\hline \hline
100 & 2.39 & 6.17 \\
\hline
150 & 2.00 & 6.27 \\
\hline
200 & 1.80 & 6.39 \\
\hline
250 & 1.69 & 6.52 \\
\hline
300 & 1.61 & 6.67 \\
\hline
400 & 1.52 & 7.03 \\
\hline
\end{tabular}
\caption{The branching ratios for the solution $V^*_{t's} V_{t'b}^{(-)}$} 
\label{tabelle1}
\end{table}

\begin{table}
\begin{tabular}{||c|c|c||}
\hline
$m_{t'} [GeV]$ & $V^*_{t's} V^{(+)}_{t'b} \times 10^{-2}$ & $Br(B_s \to \nu \bar{\nu} \gamma)$ \\
\hline \hline
100 & -10.01 & $2.23 \times 10^{-9}$ \\
\hline
150 & -8.37 & $1.79 \times 10^{-8}$ \\
\hline
200 & -7.55 & $5.22 \times 10^{-8}$ \\
\hline
250 & -7.07 & $1.13 \times 10^{-7}$ \\
\hline
300 & -6.75 & $2.10 \times 10^{-7}$ \\
\hline
400 & -6.35 & $5.58 \times 10^{-7}$ \\
\hline
\end{tabular}
\caption{The branching ratios for the solution $V^*_{t's} V_{t'b}^{(+)}$} 
\label{tabelle2}
\end{table}

\newpage

\begin{thebibliography}{1}
\bibitem{Collab} Mark II Collab., G.S. Abrams et al., \textsl{Phys. Rev. Lett.} {\bf 63} (1989) 2173; \\
                 L3 Collab., B. Advera et al., \textsl{Phys. Lett.} {\bf B231} (1989) 509; \\
                 OPAL Collab., I. Decamp et al., ibid., {\bf 231} (1989) 519; \\
                 DELPHI Collab., M.Z. Akrawy et al., ibid., {\bf 231} (1989) 539.
\bibitem{Huo} W.-J. Huo, [hep-ph/0006110]
\bibitem{Ginzburg} I. F. Ginzburg, I. P. Ivanov, A. Schiller, \textsl{Phys. Rev.} {\bf D60} (1999) 095001, [hep-ph/9802364]
\bibitem{Gunion} J.F. Gunion, D.W. McKay, H.Pois, \textsl{Phys. Lett.} {\bf B334} (1994) 339-347, [hep-ph/9406249]
\bibitem{Froggatt} C. D. Froggatt, D.J. Smith and H.B. Nielsen, \textsl{Z. Phys.} {\bf C73} (1997) 333-337, [hep-ph/9603436]
\bibitem{Beneke} M. Beneke et al., [hep-ph/0003033]
\bibitem{Huang} C.-S. Huang, W.-J. Huo and Y.-L. Wu, \textsl{Mod. Phys.} {\bf A14} (1999) 2453-2462, [hep-ph/9911203] \\
\bibitem{Chou} K.C. Chou, Y.L. Wu and Y.B. Xie, \textsl{Chinese Phys. Lett.} {\bf 1} (1984) 2
\bibitem{Aliev3} T.M. Aliev, A. \"Ozpineci and M. Savci, \textsl{Nucl. Phys.} {\bf B585} (2000) 275-289, [hep-ph/0002061]  
\bibitem{Evans} N. Evans , \textsl{Phys. Lett.} {\bf B340} (1994) 81, [hep-ph/9408308]
\bibitem{Bamert} P. Bamert and C.P. Burgess, \textsl{Z. Phys.} {\bf C66} (1995) 495, [hep-ph/9407203]
\bibitem{Inami} T. Inami et al., \textsl{ Mod. Phys. Lett.} {\bf A10} (1995) 1471
\bibitem{Masiero} A. Masiero et al., \textsl{Phys. Lett.} {\bf B355} (1995) 329, [hep-ph/9506407]
\bibitem{Novikov} V. Novikov, L.B. Okun, A.N. Rozanov et al., \textsl{ Mod.Phys. Lett.} {\bf A10} (1995) 1915
\bibitem{Erler} J. Erler, P. Langacker, \textsl{ Eur. J. Phys.} {\bf C3} (1998) 90
\bibitem{Maltoni} M. Maltoni et al., \textsl{Phys. Lett.} {\bf B476} (2000) 107-115, [hep-ph/9911535]
\bibitem{Goto} T. Goto, Y.Okada and Y. Shimizu, \textsl{Phys. Rev.} {\bf D58} (1998) 094006, [hep-ph/9804294]
\bibitem{Aliev2} T.M. Aliev, A. \"Ozpineci and M. Savci, \textsl{Phys. Lett.} {\bf B393} (1997) 143-148, [hep-ph/9610255]
\bibitem{Buras} A. J. Buras and M. M\"unz, \textsl{Phys. Rev.} {\bf D52} (1995) 186-195, [hep-ph/9501281]
\bibitem{Grinstein} B. Grinstein, M.J. Savage and M.B. Wise, \textsl{Nucl. Phys.} {\bf B319} (1989) 271
\bibitem{Huang2} C.S. Huang, Q.S. Yan, \textsl{Phys. Lett.} {\bf B442} (1998) 209-216, [hep-ph/9803366]; \\
                 C.S. Huang, W. Liao and Q.S. Yan, \textsl{Phys. Rev.} {\bf D59} (1999) 011701, [hep-ph/9803460]; \\
                 C.S. Huang and S.H. Zhu, \textsl{Phys. Rev.} {D61} (2000) 015011, [hep-ph/9905463];
\bibitem{Aliev} T. Aliev and E. Iltan, \textsl{J.Phys.} {\bf G25} (1999) 989-999, [hep-ph/9803272]
\bibitem{Buras2} A.J. Buras, [hep-ph/9806471]
\bibitem{Geng} C.Q. Geng, C.C. Lih and W.-M. Zhang, \textsl{Phys. Rev.} {\bf D57} (1998) 5697-5702, [hep-ph/9710323]
\bibitem{Dai} Y.B. Dai, C.S. Huang, H.W. Huang, \textsl{Phys. Lett.} {\bf B390} (1997) 257-262, [hep-ph/9608277]
\bibitem{Wu} Y.L. Wu, \textsl{Chinese Phys. Lett.} {\bf 16} (1999) 339, [hep-ph/9805439];
             W.-S. Hou, R. S. Willey and A. Soni (UCLA), \textsl{Phys. Rev. Lett.} {\bf 58} (1987) 1608
\end{thebibliography}
\end{document}